\documentclass[aps, prd, 12pt, showpacs, floatfix, superscriptaddress, twocolumn, nofootinbib, reprint, preprintnumbers, longbibliography]{revtex4-2}

\usepackage{multirow, amsmath, hyperref, graphicx, booktabs}
\usepackage{dcolumn}
\usepackage[dvipsnames]{xcolor}
\usepackage[final]{microtype}
\usepackage{booktabs}
\usepackage{array}

\usepackage{subcaption}
\hypersetup{
  colorlinks=true,
  citecolor=red,
  linkcolor=magenta,
  urlcolor=NavyBlue,
  allbordercolors={0 0 0},
  pdfborderstyle={/S/U/W 1}
}

\usepackage{orcidlink}

\newcommand{\IITGn}{Indian Institute of Technology Gandhinagar, Palaj Gandhinagar, Gujarat 382055, India.\vspace*{5pt}}
\newcommand{\TIFR}{Department of Astronomy \& Astrophysics, \mbox{Tata Institute of Fundamental Research}, 1, Homi Bhabha Road, Colaba, Mumbai 400005, India.\vspace*{5pt}}
\newcommand{\NIKHEF}{Nikhef, Science Park 105, 1098 XG Amsterdam, The Netherlands.}
\newcommand{\UU}{Institute for Gravitational and Subatomic Physics (GRASP), \mbox{Utrecht University}, Princetonplein 1, 3584 CC Utrecht, The Netherlands.}
\newcommand{\IWF}{Space Research Institute, Austrian Academy of Sciences, Schmiedlstrasse 6, 8042 Graz, Austria}

\begin{document}

\title{Fast and faithful interpolation of numerical relativity surrogate waveforms \\using meshfree approximation}

\author{\sc{Lalit Pathak}\orcidlink{0000-0002-9523-7945}} 
\email{lalit.pathak@iitgn.ac.in} \affiliation{\IITGn} \affiliation{\TIFR}
\author{\sc{Amit Reza}\orcidlink{0000-0001-7934-0259}} 
\email{amit.reza@oeaw.ac.at} \affiliation{\NIKHEF} \affiliation{\UU} \affiliation{\IWF}
\author{\sc{Anand S. Sengupta}\orcidlink{0000-0002-3212-0475}\vspace*{3pt}} 
\email{asengupta@iitgn.ac.in} \affiliation{\IITGn}

\begin{abstract} 
%
Several theoretical waveform models have been developed over the years to capture the gravitational wave emission from the dynamical evolution of compact binary systems of neutron stars and black holes. 
As ground-based detectors improve their sensitivity at low frequencies, the real-time computation of these waveforms can become computationally expensive, exacerbating the steep cost of rapidly reconstructing source parameters using Bayesian methods. 
This paper describes an efficient numerical algorithm for generating high-fidelity interpolated compact binary waveforms at an arbitrary point in the signal manifold by leveraging computational linear algebra techniques such as singular value decomposition and meshfree approximation. The results are presented for the time-domain \texttt{NRHybSur3dq8} inspiral-merger-ringdown (IMR) waveform model that is fine tuned to numerical relativity simulations and parameterized by the two component-masses and two aligned spins. For  demonstration, we target a specific region of the intrinsic parameter space inspired by the previously inferred parameters of the \texttt{GW200311\_115853} event -- a binary black hole system whose merger was recorded by the network of advanced-LIGO and Virgo detectors during the third observation run. We show that the meshfree interpolated waveforms can be evaluated in $\sim 2.3$ ms, which is about $\times 38$ faster than its brute-force (frequency-domain tapered) implementation in the \textsc{PyCBC} software package at a median accuracy of  $\sim \mathcal{O}(10^{-5})$. The algorithm is computationally efficient and scales favourably with an increasing number of dimensions of the parameter space. This technique may find use in rapid parameter estimation and source reconstruction studies.
\end{abstract}
\maketitle 

\section{Introduction}
\label{sec:intro}
Compact binary systems such as binary black holes or binary neutron stars are one of the most important sources for ground-based gravitational wave (GW) detectors. The GW signal emitted by such sources can be theoretically modeled in the secular inspiral or the post-merger ring-down phase by solving the Einstein's field equations. While analytical solutions are not available in the highly non-linear, so-called `merger'-domain of the evolution of these sources, the inspiral and ringdown solutions are often calibrated to a set of numerical relativity waveforms to construct complete semi-analytical waveforms covering the inspiral, merger and ringdown (IMR) phases of the evolution of such compact binary sources.
Several such waveform models have been developed both in the frequency and time-domain, such as the family of frequency-domain IMR waveforms~\cite{IMRPhenomD, IMRPhenomXAS, IMRPhenomPv2, IMRPhenomXHM, IMRPhenomXPHM}, and the EOB-family~\cite{Boh_2017, Cotesta_2020} of waveforms, that are routinely used for GW data analysis. 
More recently, numerical relativity (NR) based surrogate models have also been developed such as the \texttt{NRSur7dq2}~\cite{NRSur7dq2}, \texttt{NRSur7dq4}~\cite{NRSur7dq4}, and the \texttt{NRHybSur3dq8}~\cite{NRHybSur3dq8} models, which are among the most accurate waveform models available but are computationally expensive to generate in comparison to their frequency domain counterparts such as the \texttt{IMRPhenomXAS} and $\texttt{SEOBNRv4}\_\texttt{ROM}$ models. For reference, \texttt{NRHybSur3dq8} takes $\sim 75$~ms to generate a waveform with component masses ${m_{1, 2} = (50, 20) \, M_\odot}$ and aligned spin parameters ${\chi_{1z, 2z} = (0.05, 0.05)}$ starting at a seismic cutoff frequency of $15$~Hz. For the same set of parameters, the time to generate \texttt{IMRPhenomXAS} and $\texttt{SEOBNRv4}\_\texttt{ROM}$ is $\sim 1.5$~ms and $\sim 2.4$~ms, respectively\footnote{The timings quoted here are for the Python implementation in the \textsc{PyCBC} software package.}. 

Our research is inspired by a suggestion by Verma et al.~\cite{NRHybSur3dq8}, where the authors table the idea of speeding up the evaluation of the \texttt{NRHybSur3dq8} time-domain surrogate waveforms by creating a faster frequency-domain variant. 
In the past, fast frequency-domain surrogates using the technique of model order reduction have been applied to the SEOBNRv4~\cite{Boh_2017} waveform model leading to a significant reduction in computational complexity of evaluating these waveforms. Such an approach could potentially accelerate parameter estimation (PE) for compact binary sources using these accurate waveform models.

There have been several attempts in the past to construct surrogate models of other waveform models in the past. 
Cannon et al.~\cite{cannon2012interpolating} constructed the approximate non-spinning IMR waveforms~\cite{Ajith_2011} parameterized by only the two-component masses - by first projecting these waveforms over a  set of singular value of decomposition (SVD) basis vectors, followed by a grid-based two-dimensional Chebyshev interpolation of the SVD coefficients.
Chua et al.~\cite{PhysRevLett.122.211101} used artificial neural networks (ANNs) to generate a four-dimensional reduced order model (ROM) of the $2.5$ post-Newtonian (PN) frequency domain \texttt{TaylorF2} waveform model~\cite{PhysRevD.79.104023} parameterized by the two-component masses and the two aligned spins. In their approach, the waveforms are represented as the weighted sums over the reduced basis vectors. The ANNs are then trained to accurately map the GW source parameters into the basis coefficients. 
Other studies related to building NR surrogate models utilized ANNs~\cite{Khan_2021}, Gaussian Process Regression (GPR)~\cite{Williams_2020}, and deep learning~\cite{Lee_2021} architectures.

\begin{figure*}[!hbt]
\centering
\includegraphics[width=\textwidth]{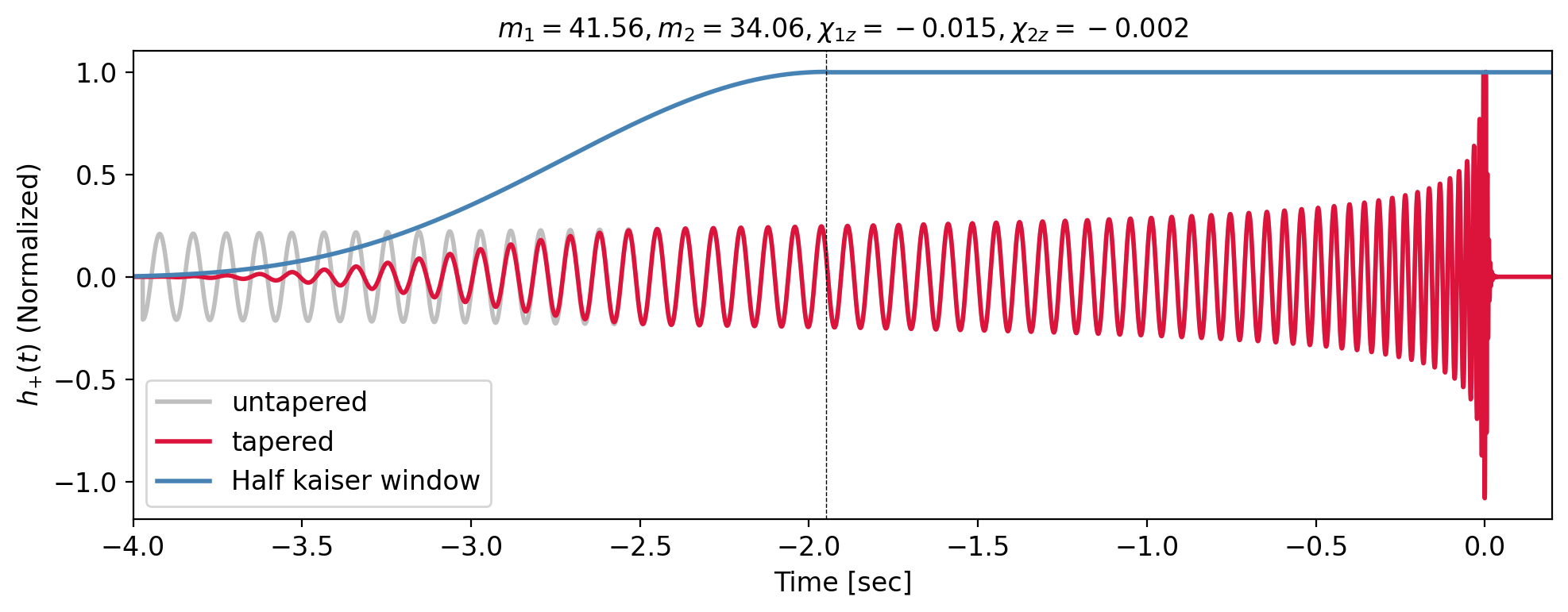}
\caption{The \texttt{NRHybSur3dq8} is generated at a frequency ($f_{start} = 10$ Hz) lower than the seismic cutoff frequency ($f_{low} = 15$ Hz), which leads to a longer duration waveform. It is generated for a simulated BBH event whose intrinsic parameters are shown in the title. A time-domain tapering using a half-Kaiser window is performed on the longer waveform to smoothly attenuate the waveform amplitude to zero, that helps in alleviating the Gibbs phenomenon arising due to a jump discontinuity while converting the time-domain waveform to a frequency-domain waveform. The black dashed vertical line represents the epoch at which the amplitude of the tapered and un-tapered waveform becomes equal (or equivalently, the value of the half-Kaiser window becomes $1$). The sampling frequency is set at $2048$ Hz. The waveform amplitude ($y$-axis) is normalized by the maximum value of the strain.}
\label{fig:wf_tap_vs_untap_td}
\end{figure*} 

In this work, we aim to significantly reduce the computational cost of generating frequency-domain \texttt{NRHybSur3dq8} waveforms by using a combination of the SVD-decomposition and meshfree approximation. We focus on illustrating our approach by concentrating on a region of the four-dimensional (two component masses and two aligned spins) parameter space around the inferred source parameters of the  \texttt{GW200311\_115853}~\cite{PhysRevX.13.011048} event detected during the third observing run of the LVK collaboration~\cite{Abbott_2020_Prospects}. We construct meshfree interpolants for both the amplitude and phase of these waveforms separately. These interpolants are then used to rapidly evaluate the amplitude and phase of the waveform at any arbitrary query points in the chosen parameter space. These interpolated waveforms can be evaluated in $\sim 2.3$ ms in comparison to $89$ ms ($\sim 38 \times $ faster) with the standard (frequency-domain tapered) implementation of time-domain \texttt{NRHybSur3dq8} in \textsc{PyCBC}~\cite{usman2016pycbc} with a median error of  ${\sim \mathcal{O}(10^{-5})}$. 
Further, we also performed a Bayesian PE study of a simulated \texttt{GW200311\_115853} like event coherently injected in Gaussian noise to mimic data from the network of LIGO and Virgo detectors. The PE runs were carried out using the meshfree interpolated waveforms and took ${\sim 16.4 \, \text{minutes}}$ to complete on a $64$ CPU-cores setup and the posterior distributions over the source parameters were found to be broadly consistent with their injected values. In comparison, with a bruteforce frequency domain implementation of tapered time-domain \texttt{NRHybSur3dq8} the posteriors were generated in \mbox{$\sim 5$ h and $50$ min} (on $64$ CPU-cores) underscoring the gains achieved with the frequency-domain meshfree waveforms.

The rest of the paper is organized as follows: Section~\ref{sec:wave_model} introduces the \texttt{NRHybSur3dq8} waveform model and the preprocessing required before we build meshfree interpolants. Section~\ref{sec:interp_generation}
explains an iterative strategy to find a suitable set of basis vectors using SVD to span the space of amplitude and phase and construct the meshfree interpolants of the resulting SVD coefficients. In Section~\ref{sec:choice_of_rbf_kernel}, we describe various RBF kernels that can be used to generate meshfree interpolants. Section~\ref{sec:results} demonstrates the results of a PE performed on a simulated BBH event using meshfree interpolants.  Finally, we summarize the results in Section~\ref{sec:conclusion} and discuss the limitations of the current implementation and suggest some ideas for overcoming this limitation in follow-up studies.

\section{NR Waveform model}
\label{sec:wave_model}
\texttt{NRHybSur3dq8} is a time-domain, aligned spin surrogate model for ``hybridized" non-precessing NR waveform, which is valid for stellar compact binaries with total masses as low as $2.5\, M_{\odot}$. In this context, the term ``hybridized" refers to a combined waveform incorporating both a post-Newtonian (PN) and effective one-body (EOB) waveform during the early times, attached smoothly to numerical relativity (NR) waveform at late times. The training of this model involves hybridized waveforms derived from $104$ numerical relativity (NR) waveforms, supporting mass ratios $q \leq 8$, and spin values $\chi_{1z}$; $\chi_{2z} \leq 0.8$. Here, $\chi_{2z}$$(\chi_{2z})$ represents the spin of the heavier (lighter) black hole in alignment with the direction of orbital angular momentum. The waveform model also supports spin-weighted spherical harmonic modes with $l \leq 4$ and $(5, 5)$ mode but excludes the $(4, 0)$ or $(4, 1)$ modes. In this work, we specifically focus on the dominant, commonly referred to as the "quadrupole" mode $(2, 2)$ present in this waveform model. 
The GW polarizations, denoted as ``plus" $(h_{+})$ and ``cross" $(h_{\times})$, can be concisely represented as a unified complex time series, denoted as {${h = h_{+} - j\,h_{\times}}$}. This complex time series can be expressed as a linear combination of spin-weighted harmonic modes $h_{lm}$~\cite{Newman:1966ub, Goldberg:1966uu}. Consequently, gravitational waves propagating along any direction $(\iota, \phi_0)$ in the binary source's frame can be expanded in spin weighted spherical harmonic functions $^s Y_{lm}$ with spin weight ${s=-2}$ as:
\begin{equation}
    h(t, \iota, \phi_0) = \sum_{l=2}^{\infty}\sum_{m=-l}^{l}\, h_{lm}(t) \, ^{-2}Y_{lm}(\iota, \phi_0),
    \label{eq:wave_spheric_modes}
\end{equation}
%
where, $\iota$ represents the inclination angle between the orbital angular momentum of the binary and the line of sight to the detector, and $\phi_0$ corresponds to the initial binary phase. The corresponding frequency domain waveform for the dominant mode can be written as the following:
\begin{figure*}[!hbt]
\begin{subfigure}{0.49\linewidth}
    \centering
    \includegraphics[width=\linewidth]{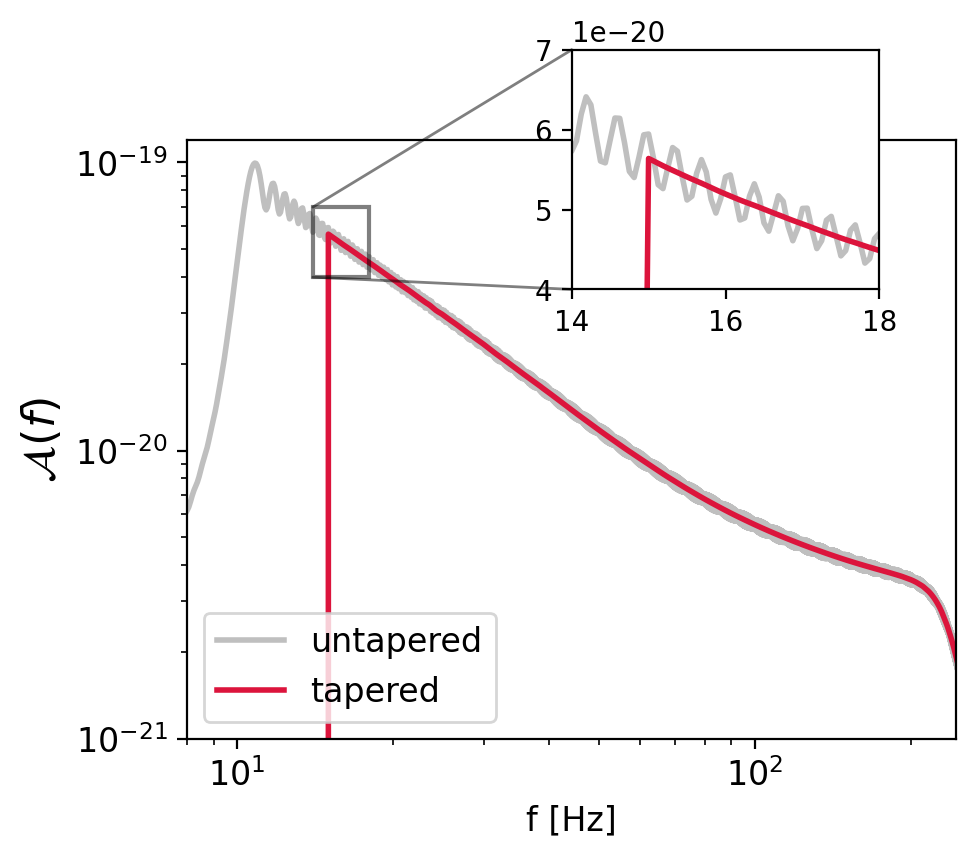}
    \caption{}
    \label{fig:wf_tap_vs_untap_fd}
\end{subfigure}\hfill
\begin{subfigure}{0.49\linewidth}
    \centering
    \includegraphics[width=\linewidth]{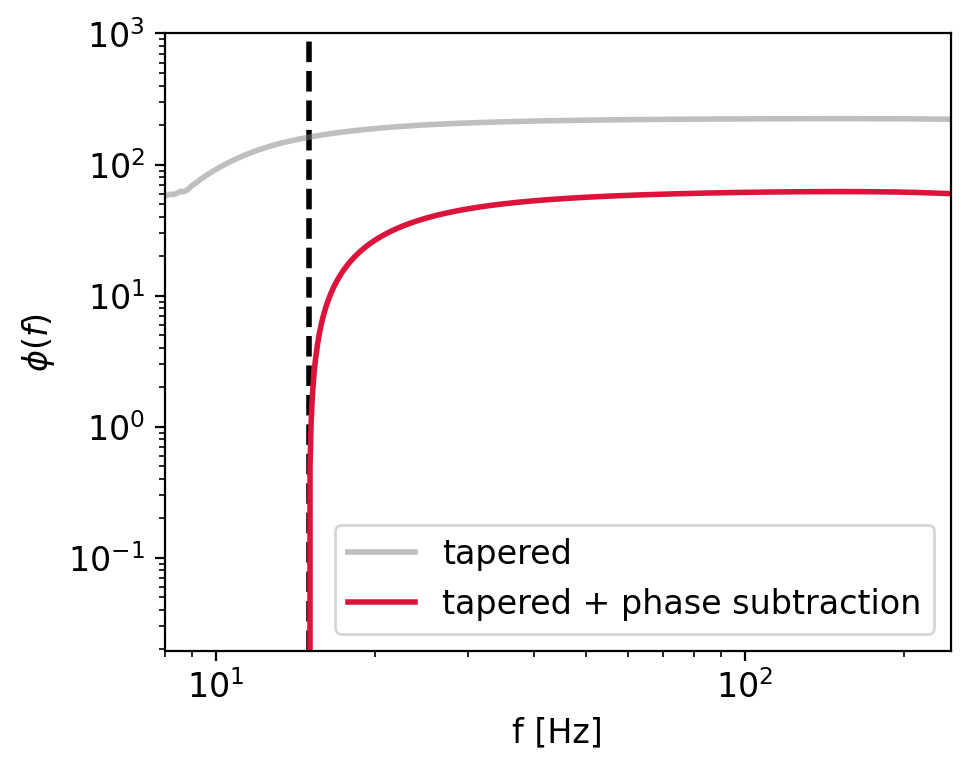}
    \caption{}
    \label{fig:wf_tap_vs_untap_fd_phase}
\end{subfigure}
\caption{\textit{left}: The amplitude of the untapered waveform (\textit{solid grey}) is plotted with that of the tapered (\textit{solid red}) waveform. Note the ringing artifacts appearing due to the Gibbs phenomenon in the untapered waveform. It is almost negligible in the amplitude of the tapered waveform. \textit{right}: After converting the tapered waveform into the frequency domain, there is some accumulated phase below $15$ Hz, which we subtract from the total phase so that the phase of the frequency domain waveform is zero at ${f_{low} = 15}$ Hz. The black-dashed vertical line represents ${f_{low} = 15}$ Hz.}
\end{figure*}
\begin{equation}
    \tilde{h}_{+/\times} = \mathcal{A}_{+/\times}(f)\ \exp[j\,\psi_{+/\times}(f)]
    \label{eq:frequency_domain_waveform}
\end{equation}
where $\mathcal{A}_{+/\times}(f)$ and $\psi_{+/\times}(f)$ are amplitude and phase as a function of frequency, respectively and $j$ represents the complex number $j = \sqrt{-1}$. Since \texttt{NRHybSur3dq8} is an aligned-spin waveform model~\cite{PhysRevD.49.1707, PhysRevD.59.124016, Faye_2012, PhysRevLett.74.3515, khan2016frequency, husa2016frequency}, the relation ${\tilde{h}_{\times} \propto -j\: \tilde{h}_{+}}$ holds and therefore we only consider $\tilde{h}_{+}$ and the cross-polarization ($\tilde{h}_{\times}(f)$) can be calculated from the above relation. From now onwards, we will drop the subscript `+/$\times$' from $\mathcal{A}_{+/\times}$ and $\psi_{+/\times}$ and simply denote the amplitude and phase by $\mathcal{A}$ and $\psi$ which correspond to plus polarization.
\begin{figure*}[!hbt]
\begin{subfigure}{0.49\linewidth}
    \centering
    \includegraphics[width=\linewidth]{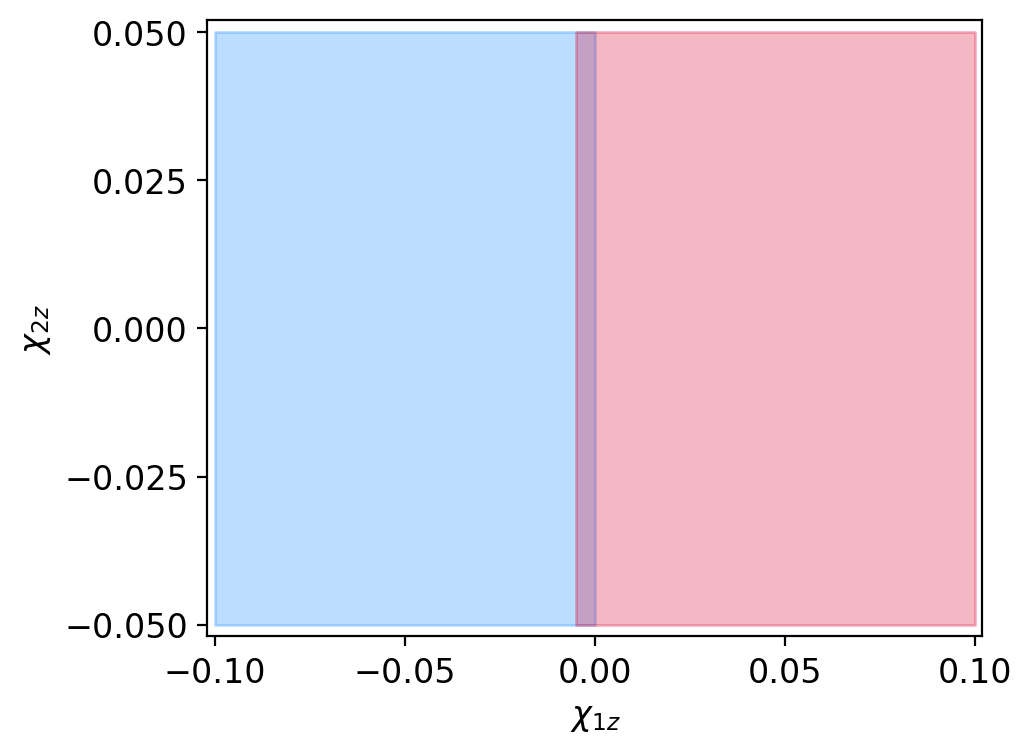}
    \caption{}
    \label{fig:overlap_patch}
\end{subfigure}\hfill
\begin{subfigure}{0.49\linewidth}
    \centering
    \includegraphics[width=\linewidth, height=0.75\linewidth]{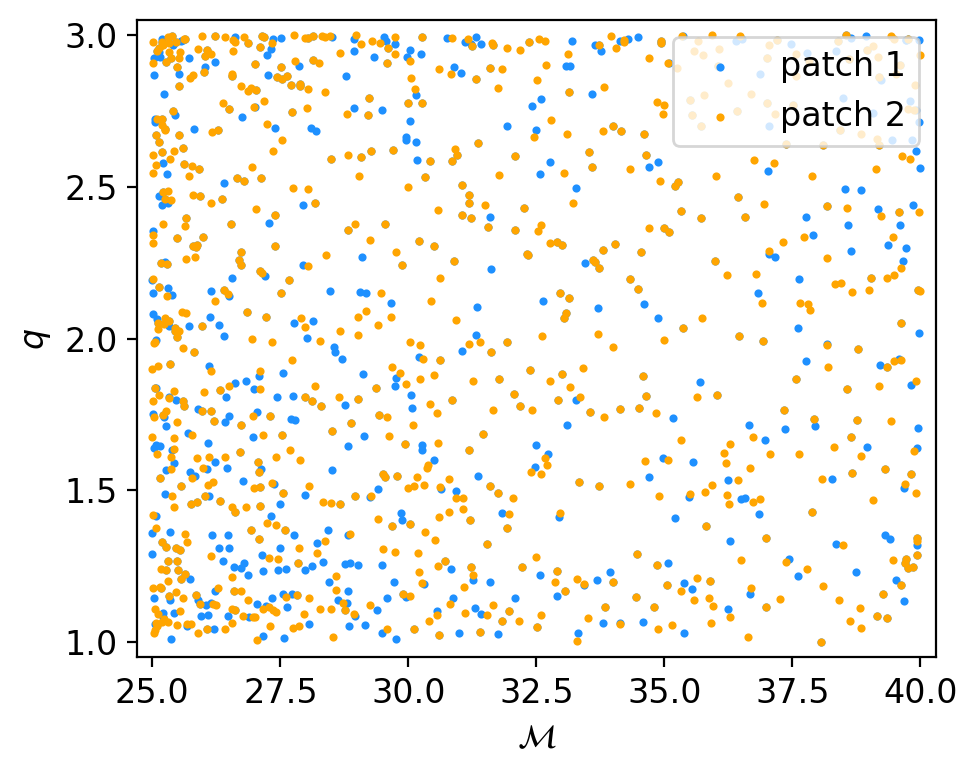}
    \caption{}
    \label{fig:final_nodes_both_patches}
\end{subfigure}
\caption{\textit{left}: We choose two overlapping patches over $\chi_{1z}$ to cover the desired range in $\chi_{1z}$. A similar procedure can be performed to extend the range of the other parameters (say ${\mathcal{M}, \, q}$ etc.). \textit{right}: Final set of nodes after iterative SVD process terminates for both the patches. The nodes found using the ``Iterative SVD'' procedure tend to cluster near the lower mass boundaries.}
\end{figure*}

As a pre-processing step, we first calculate the duration of the waveform starting at the seismic cutoff frequency ($f_{low}$). Subsequently, the starting frequency ($f_{start}$) of the waveform generation is decreased until we reach twice the duration of the desired waveform. Then the time-domain \texttt{NRHybSur3dq8} waveform is generated starting at $f_{start}$ followed by a time-domain tapering to smoothly decrease the amplitude to zero (see Fig.~\ref{fig:wf_tap_vs_untap_td}). The length of the tapering window is taken as a fraction ($0.8$ in this case) of the difference between the new and desired duration of the waveform. The longer duration of the waveform makes sure that we do not lose any portion of the waveform in the frequency band to be used in the PE while tapering. We use a Kaiser window~\cite{kuo1966system} (as implemented in \textsc{}{PyCBC}) for time-domain tapering, which can be expressed as
\begin{equation}
    w(n) = I_0\left(\beta \, \sqrt{1 - \frac{4n^2}{(M-1)^2}}\right)/I_0(\beta)
    \label{eq:kaiser_window}
\end{equation}
with $n \leq |\frac{M-1}{2}|$ and where $I_0$ is the modified zeroth-order Bessel function, $M$ is the number of points in the output window, and $\beta$ is the shape parameter, which determines the trade-off between main-lobe width and side lobe level in the Fourier response of the window. As the parameter $\beta$ becomes large, the main lobe width increases while the side lobe level decreases. For this analysis, we use a half-Kaiser window for tapering (see Fig.~\ref{fig:wf_tap_vs_untap_td}). The tapering helps in reducing the Gibbs phenomenon while converting the tapered waveform in the frequency domain (see Fig.~\ref{fig:wf_tap_vs_untap_fd}). Additionally, we also subtract the phase accumulated up to $f_{low}$ from the total phase of the frequency domain waveform (tapered) to make sure the phase at $f_{low}$ is zero, which significantly helps in the interpolation of the SVD coefficients corresponding to the phase (see Fig.~\ref{fig:wf_tap_vs_untap_fd_phase}). 
%

In the next section, we lay down an iterative strategy to find suitable basis vectors to span the space of amplitude and phase using SVD. Subsequently, we fit the resulting SVD coefficients for amplitude and phase separately using a linear combination of RBFs and monomials. Since the amplitude and phase are smoothly varying functions of frequency (see Fig.~\ref{fig:wf_tap_vs_untap_fd} and Fig.~\ref{fig:wf_tap_vs_untap_fd_phase}), the corresponding SVD coefficients are expected to exhibit smooth variation over the intrinsic parameter space as well and, therefore, suitable for interpolation. Combining the interpolated coefficients at the arbitrary query points within the interpolating region with the corresponding basis vectors gives the interpolated amplitude and phase and, hence, the interpolated waveform.

\section{Generation of meshfree interpolants}
\label{sec:interp_generation}
To construct the interpolants for evaluating the meshfree waveforms, the initial step involves choosing a patch of intrinsic parameter space for which interpolants are to be built. The dimensionality of the patch depends on the dimensionality of the intrinsic parameter space over which interpolation is to be performed. Rather than opting for a single patch with large boundaries, a more effective approach (as suggested in~\cite{P_rrer_2014}) is to create multiple overlapping patches with smaller boundaries to comprehensively encompass the desired parameter space. For each patch, mesh-free interpolants are independently constructed, enhancing the accuracy of individual interpolants. In this analysis, the selection of patches is inspired by a simulated BBH event with \texttt{GW200311\_115853}-like parameters. We use \texttt{NRHybSur3dq8}, an aligned spin numerical relativity (NR) surrogate waveform model, which also includes support for subdominant modes. As a proof of principle study, this analysis focuses only on the leading-order mode; however, the procedure presented here can be extended to incorporate higher-order modes. The procedure to generate interpolants comprises the following steps:

\subsection{Selection of patch(es)}
This task involves identifying the patch (or patches) within the parameter space for the construction of waveform interpolants. A similar approach was taken by Morisaki et al. in their work~\cite{morisaki2023rapid} for constructing ROM bases corresponding to different patches of parameter space. While it could be any segment of the pertinent parameter space in general, for this study, we opt for a patch encompassing the injected values of parameters associated with a simulated BBH event as described earlier. 
The number of patches should be selected based on the specific research problem, whether it involves fast parameter estimation (PE) or waveform generation. Our goal is to rapidly obtain posterior distributions, so the number of patches depends on the injection parameters for a simulated signal. For a real GW event, the injection parameters are replaced by the best-matched template. Fewer patches are needed for a PE run because the prior range of parameters is small. The primary criterion is that the patches must encompass the injection parameters, with their width chosen based on expected estimation errors. If the patches do not include the injection parameters, biases in parameter estimation are likely to occur, as in any Bayesian PE analysis. Additionally, the patch width should be selected to generate accurate interpolants, which can be achieved by using multiple patches in the intrinsic parameter space with smaller widths. For waveform generation, a relatively large number of patches is necessary to cover a larger parameter space.

For this work, we define two four-dimensional hyper-rectangular patches with the following ranges: ${\mathcal{M} \in [25, \, 40]}$, ${q\in [1, \, 3]}$, ${\chi_{1z} \in [-0.1, \, 0.1]}$, and ${\chi_{2z} \in [-0.05, \, 0.05]}$. Both patches share common ranges for $\mathcal{M}$, $q$, and $\chi_{2z}$. However, for the first patch, ${\chi_{1z} \in [-0.1, \, 0]}$, and for the second patch, ${\chi_{1z} \in [-0.005, \, 0.1]}$. Note that these two patches overlap in the $\chi_{1z}$ dimension (see Fig.~\ref{fig:overlap_patch}).

\subsection{Placing grid of RBF nodes using `Iterative SVD' method }
\label{subsec:itr_svd}
This step starts by randomly spraying a minimum (initial) number of nodes across the four-dimensional sample space to start the algorithm and generating frequency domain waveforms (as specified in Sec.~\ref{sec:wave_model}) at these specified nodes. The minimum number of nodes can be calculated using {${N = \binom{\nu + \mathcal{D}}{\nu}}$} where $\nu$ is the degree of the monomial (see Eq.~\eqref{eq:amp_phi_rbfcoeff}) and $\mathcal{D}$ is the dimensionality of the intrinsic parameter space~\cite{doi:10.1142/6437, pathak2022rapid}. Subsequently, $\mathcal{A}$ and $\psi$ are extracted from $\tilde{h}_{+}(f)$ for each of the $N$ waveforms. The obtained amplitudes ($\mathcal{A}$) and phases ($\psi$) are then stacked in a row-wise fashion to construct two matrices, $\mathbf{A}$ and $\mathbf{\Psi}$. These matrices are subjected to singular value decomposition (SVD), which, in turn, provides the relevant basis for spanning the space defined by these amplitudes and phases.
\begin{equation}
    \mathcal{A} = \sum_{\mu = 1}^{N} C^{\mathcal{A}}_{\mu} \; \vec u^{\,\mathcal{A}}_{\mu} \quad\text{and}\quad \psi = \sum_{\mu = 1}^{N} C^{\psi}_{\mu} \; \vec u^{\,\psi}_{\mu}
    \label{eq:svd_amp_phase}
\end{equation}
\begin{figure}[!hbt]
    \centering
    \includegraphics[width=1\linewidth]{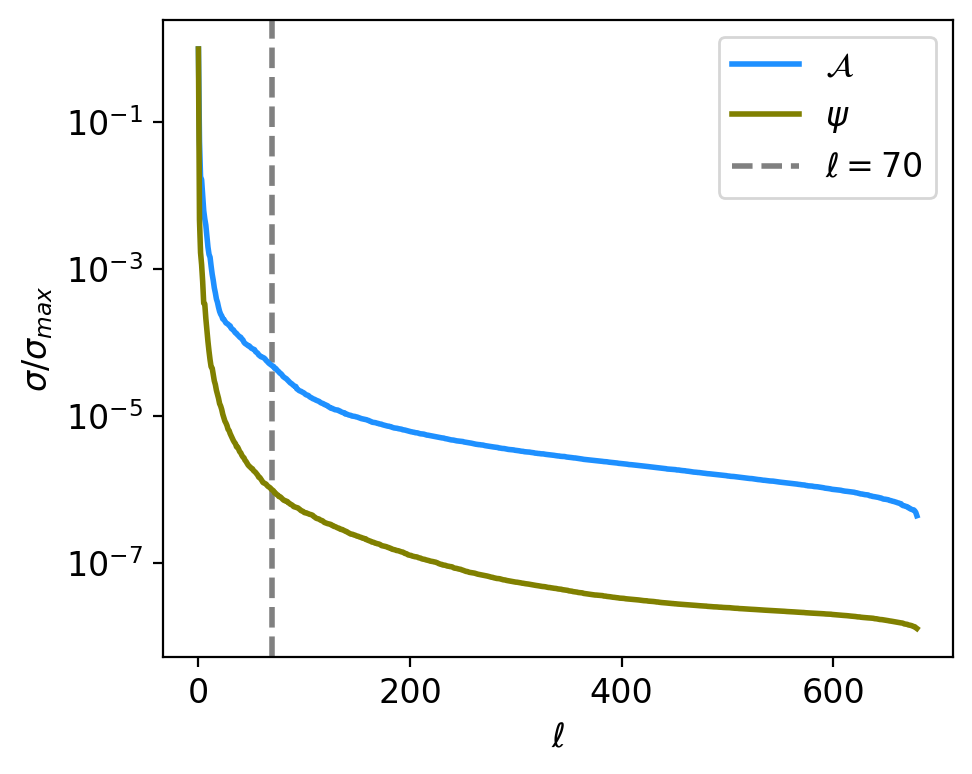}
    \caption{Spectrum of singular values corresponding to amplitude $\mathcal{A}$ and phase $\varphi$ (see Eq.~\eqref{eq:svd_amp_phase}), normalized to the maximum value. Note that singular values (relative to the maximum) drop to $\sim \mathcal{O}(10^{-6})$ within the top-$70$ basis vectors.}
    \label{fig:sing_vals}
\end{figure}

Since by construction, the basis vectors $(\vec u_{\mu}^{\mathcal{A}}, \vec u_{\mu}^{\,\psi})$ are in the decreasing order of importance, it turns out that we need to retain only top few basis vectors (say upto $\ell$) to accurately reconstruct amplitude and phase as illustrated by the singular values spectrum in Fig.~\ref{fig:sing_vals}. Consequently, we limit the construction of interpolants to the SVD coefficients $C^{\mathcal{A}}_{\mu}$ and $C^{\psi}_{\mu}$ corresponding to top-$\ell$ basis vectors. Later on, we can choose to retain an even lesser number of basis vectors while reconstructing amplitude and phase at arbitrary query points. These coefficients, denoted as $C^{\mathcal{A}}_{\mu}$ and $C^{\psi}_{\mu}$, exhibit smooth variations as functions of $\vec \lambda$, where ${\vec \lambda \equiv \{\mathcal{M}, q, \chi_{1z}, \chi_{2z}\}}$ represents the interpolating parameter space. These coefficients can be written as a linear combination of RBFs ($\phi$) and monomials ($p$) of specified degree as follows~\cite{doi:10.1142/6437}:
\begin{subequations}
    \begin{align}
    C^{\mathcal{A}\, (q)}_{\mu} &= \sum_{n=1}^N\, a^{\mathcal{A}}_{n}\, \phi(\|\vec \lambda^q - \vec \lambda^{n}\|_2) + \sum_{k = 1}^{M}\, b^{\mathcal{A}}_{k}\, p_k(\vec \lambda^q) \\
    C^{\psi\, (q)}_{\mu} &= \sum_{n=1}^N\, a^{\psi}_{n}\, \phi(\|\vec \lambda^q - \vec \lambda^{n}\|_2) + \sum_{k = 1}^{M}\, b^{\psi}_{k}\, p_k(\vec \lambda^q) 
    \label{eq:amp_phi_rbfcoeff}
    \end{align}
\end{subequations}
The addition of these monomial terms enhances the accuracy of RBF interpolation, as shown in~\cite{2016JCoPh}. From the above equations, it is clear that we need $(N+M)$ equations to uniquely solve for the $(N+M)$ RBF and monomial coefficients ($a^{\mathcal{A}/\psi}$ and $b^{\mathcal{A}/\psi}$). The SVD coefficients ($C^{\mathcal{A}\, (q)}_{\mu}$ and $C^{\psi\, (q)}_{\mu}$) are known at the $N$ RBF nodes. $M$ additional conditions are imposed by demanding  ${\sum_{k=1}^M a^{\mathcal{A}/\psi}_k p_k(\vec \lambda^q) = 0}$ which leads to a system of $(N+M)$ linear equations that can be used to uniquely determine the RBF and monomial coefficients.
Using these RBF coefficients, we can quickly evaluate the SVD coefficients $C^{\mathcal{A}\, (q)}_{\mu}$ and $C^{\psi\, (q)}_{\mu}$ (corresponding to the $\mu^{th}$ SVD basis vector) at an arbitrary query point within the interpolating parameter space and then combine it with corresponding basis vectors ($\vec u^{\,\mathcal{A}}$ and $\vec u^{\,\psi}$) to get back the interpolated amplitude $\mathcal{A}$ and phase $\psi$. Then, we combine both amplitude and phase in accordance with Eq.~\eqref{eq:frequency_domain_waveform} to get the interpolated frequency domain waveform.  
Once we have the interpolants ready, a set of $10^{3}$ query points is generated within the interpolating patch, and corresponding true frequency domain waveforms and meshfree interpolated waveforms are generated at these query points. Subsequently, the mismatches between the true and interpolated waveforms are calculated. A specified number of query points ($10$ for this work) exhibiting the worst mismatch are added back into the original set of initial nodes. The process then reverts to Subsection~\ref{subsec:itr_svd} and continues until either a limit of maximum nodes is reached or the maximum mismatch at the current iteration falls below a predefined threshold (e.g., ${\leq 10^{-4}}$) on maximum mismatch. Note that the mismatches are calculated between meshfree waveforms and waveforms first generated in time-domain at $f_{start} = 10$ Hz, then tapered and converted into frequency domain followed by phase subtraction as mentioned in Section~\ref{sec:wave_model}. We define the match between the two waveforms $h_1$ and $h_2$ as
\begin{equation}
    \mathcal{O}(h_1, h_2) = \frac{\langle h_1 \mid h_2\rangle}{\sqrt{\langle h_1 \mid h_1\rangle\langle h_2 \mid h_2\rangle}}
    \label{eq:match}
\end{equation}
where,
\begin{equation}
    \langle h_1 \mid h_2\rangle = \int_{f_{\text{low}}}^{f_{\text{high}}} \frac{\tilde{h}_{1}(f)\tilde{h}^{\ast}_{2}(f)}{S_n(f)} \, df   
\end{equation}
which defines a noise-weighted inner product where $S_n$ is the one-sided noise power spectral density and the asterisk ($*$) denotes the complex conjugation. Finally, the mismatch is defined as: 
\begin{equation}
\text{Mismatch} = 1 - \mathcal{O}(h_1, h_2)
\end{equation}
where the match $\mathcal{O}(h_1, h_2)$ is maximized over time and phase shift between the two waveforms.

We call this process ``Iterative SVD'' because it progressively refines the meshfree waveform approximation by reinserting the query points with the worst mismatch back into the initial set of nodes, followed by evaluation of the true frequency domain waveforms at the appended set of nodes. 

Currently, with each iteration of the ``iterative SVD" node placement algorithm, as additional RBF nodes are introduced, the data matrices $\mathbf{A}$ and $\mathbf{\Psi}$ are expanded by incorporating more rows (one for each newly added node) followed by a Singular Value Decomposition (SVD) conducted on these expanded data matrices.
This approach is practical for handling the decomposition of the small-sized data matrices. 
However, one can update the basis vectors and singular values at each iteration with more sophisticated algorithms ~\cite{BRAND200620, Stange2008OnTE}  
that update the left and right subspace and singular values from the existing SVD decomposition of a dense matrix as new rows are added to it. 
For updating the subspaces and singular values, Brand et al. \cite{BRAND200620} developed an algorithm for computing a thin SVD of a data matrix of size (${m \times n}$) in a single pass with linear time complexity $\mathcal{O}(mnr)$, where $r \leq \sqrt{\text{min} (m, n)}$. The authors proposed fast and memory-efficient sequential algorithms for tracking the singular values and subspaces, initialized with a general identity included in the existing decomposing of the data matrix. Adding a new column would modify the identity representation and provide \mbox{rank one} updates via the modified Gram–Schmidt algorithm. 

Note that if the required number of RBF nodes is small, changing the strategy for computation of the subspace and coefficients using the ``updated SVD'' approach would not significantly reduce the overall computational cost. 
On the other hand, for a large number of nodes, the incorporation of such sophisticated linear algebra algorithms in our proposed framework would be beneficial. We will explore this possibility in detail in future work.

\begin{figure}[!hbt]
    \centering
    \includegraphics[width=1\linewidth]{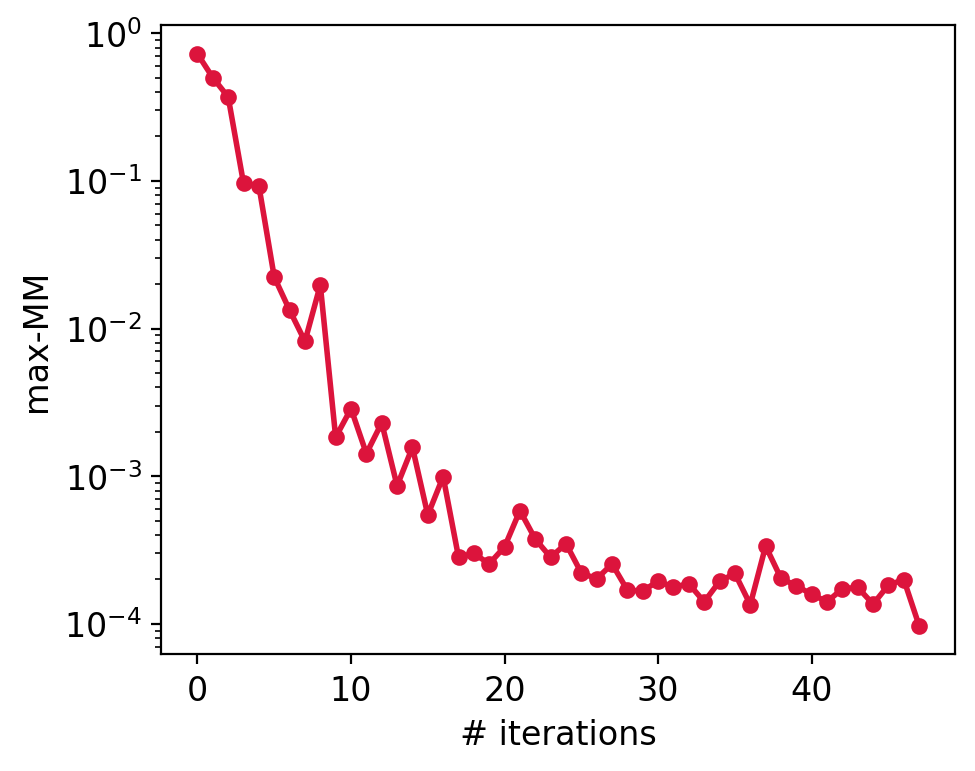}
    \caption{Mismatch between the worst approximated meshfree waveform and true waveform at each iteration. As evident from the figure, with each iteration, the meshfree approximation becomes better.}
    \label{fig:max_mismatch_at_eachitrn}
\end{figure}

Next, the SVD is performed for the new amplitude and phase matrices. This is a similar approach that was used in previous studies for building reduced-order models (ROMs) of a number of waveform approximants~\cite{canizares2013gravitational, canizares2015accelerated, Morisaki_2020, morisaki2023rapid, effroqs23}. In each iteration, SVD is employed to identify improved basis vectors for $\mathcal{A}$ and $\psi$. As shown in Fig.~\ref{fig:max_mismatch_at_eachitrn}, the maximum mismatch calculated between the worst meshfree waveform and the corresponding true waveform decreases as we add the query points corresponding to the worst meshfree waveforms at each iteration 
into the set of nodes used to generate the interpolants. The calculation of mismatches assumes a {\texttt{aLIGOZeroDetHighPower}}~\cite{aLIGO_ZDHP} Power Spectral Density (PSD) as implemented in \textsc{PyCBC}. We repeat the same ``Iterative SVD'' procedure for the other interpolating patches (if needed). For the analysis in this paper, we choose two patches (see Section~\ref{sec:results}). As evident from Fig.~\ref{fig:final_nodes_both_patches}, RBF nodes (for both patches) found by the ``Iterative SVD'' tend to cluster more towards the low-mass boundaries implying that the points near the lower mass boundaries contribute the most in terms of the suitable basis vectors for each patch. Similar patterns are also seen in the work focusing on building a surrogate of PN waveforms~\cite{NRHybSur3dq8} to find the desired training dataset of the parameters and building ROM using ANNs~\cite{chua2019reduced}, using a greedy strategy. 

Note that initially, the maximum mismatch at each iteration is decreasing rapidly. After $\sim 25$ iterations, the rate of decrease in the mismatch reduces despite adding $10$ worst points at each iteration. It implies that it is close to having the minimum number of nodes required to satisfy the maximum mismatch threshold criteria mentioned earlier. After $\sim 48$ iterations, the maximum mismatch falls below $10^{-4}$, and algorithm stops. However, to confirm whether it has indeed crossed this criterion (since there are fluctuations), we can test the accuracy of interpolants again by spraying a different set of random query points within the interpolating parameter space and confirm that the maximum mismatch is still less than $10^{-4}$. Otherwise, we can continue the algorithm until this threshold criterion is met by a few sets of random query points. 
\section{Choice of RBF kernel}
\label{sec:choice_of_rbf_kernel}
\begin{figure}[hbt]
    \centering
    \includegraphics[width=\linewidth]{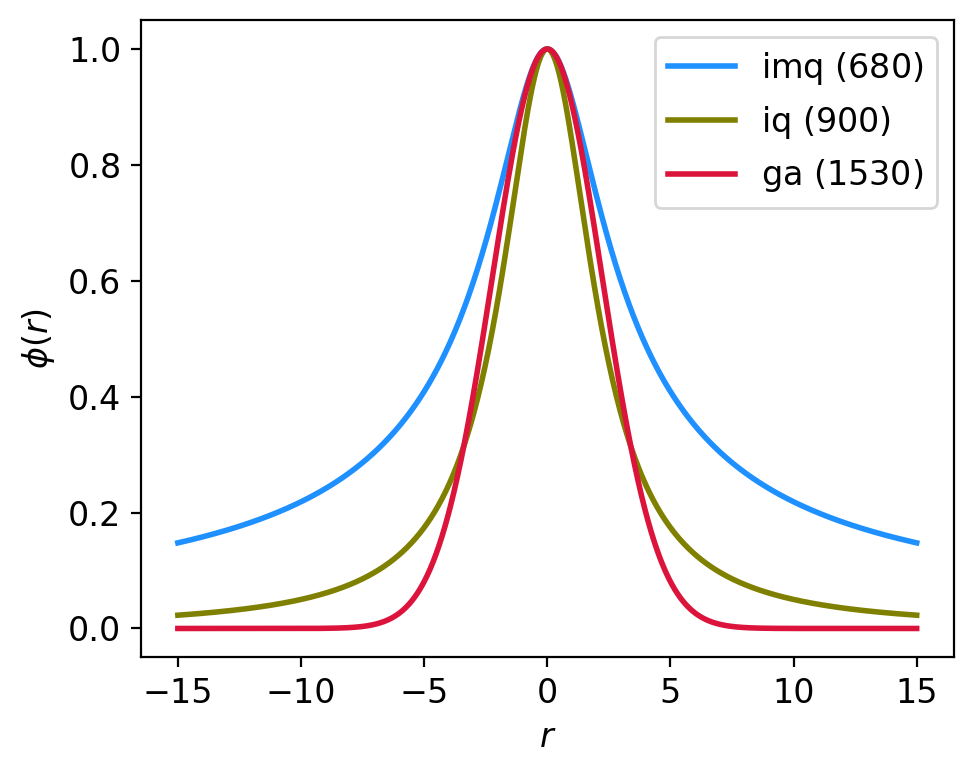}
    \caption{Profiles of different RBF kernels. In the legend, numbers in the bracket are the final number of nodes required (for patch $1$) to fall below the maximum mismatch threshold of $10^{-4}$ in the iterative SVD process. For the `imq' kernel, we required the least number of nodes to achieve the threshold.}
    \label{fig:diff_rbf_kernels}
\end{figure}

There are different RBF kernels that can be used as a basis (see Eq.~\eqref{eq:amp_phi_rbfcoeff}) for meshfree approximation. In this work, we tried the following three different RBF kernels (see Table~\ref{tab:rbf_kernels}): (i) Inverse multiquadric (`imq'), (ii) Inverse Quadric (`iq'), and (iii) Gaussian (`ga'). As mentioned earlier in sec.~\ref{sec:interp_generation}, to terminate the iterative SVD, we set a threshold of either a maximum number of nodes (3000 here) or a maximum mismatch of $10^{-4}$ or less at a given iteration. 
\begin{table}[!hbt]
    \renewcommand{\arraystretch}{1.75}
    \begin{ruledtabular}
        \begin{tabular}{l c c c}
            RBF kernel type & $\phi(r)$ & \multicolumn{2}{c}{$\epsilon$} \\ \cmidrule[0.8pt](rr{0.95em}){3-4}
            & & Patch 1 & Patch 2 \\ \hline
            inverse multiquadric (`imq') & $1/\sqrt{(1 + (\epsilon\,r)^2)}$ &  0.4016 & 0.4010 \\
            Gaussian (`ga')              &  $\exp[-(\epsilon\, r)^2]$ & 0.3172 & 0.2909 \\
            inverse quadric (`iq')       &  $1/(1 + (\epsilon\,r)^2)$& 0.4464 & 0.4378 \\ 
        \end{tabular}
    \caption{Different RBF kernels}
    \label{tab:rbf_kernels}
    \end{ruledtabular}    
\end{table}
The interpolant generation is stopped whenever any of the above-stated conditions is satisfied.  In the case of `ga'(`iq') kernels, the final number of nodes for patch $1$ and patch $2$ was $1530(900)$ and $1620(590)$, respectively, when the maximum mismatch at an iteration fell below the threshold of $10^{-4}$. However, in the case of `imq' RBF kernel, the final number of nodes for patch one was $680$, while for the second patch, it was $770$ for the same threshold. A natural question arises: what is so special about the `imq'? A possible explanation is the spread of these RBF kernels. As shown in Fig.~\ref{fig:diff_rbf_kernels}, the `ga', `imq', and `iq' kernels have similar profiles near the center ($r=0$). However, as we go away from the center, the `ga' and `iq' kernels fall very rapidly in comparison to the `imq' kernel. It implies that the `imq' kernel has a significant overlap with the `imq' kernel centered at the other nearby nodes (centers) in comparison to the other two RBF kernels, and therefore, it explains the lesser number of nodes required to fall below the maximum mismatch threshold at a given iteration for `imq' in comparison to a relatively higher number of nodes for the other RBF kernels.
For the next section, we choose `imq' as the RBF kernel for generating meshfree interpolants. 

\section{Results}
\label{sec:results}
As previously mentioned, there are two overlapping patches in $\chi_{1z}$, and we independently construct interpolants for each of the patches. In the first patch where ${\chi_{1z} \in [-0.1, \,0]}$, we initiate with ${N = \binom{\nu + d}{\nu}}$ nodes, with $\nu$ representing the degree of the monomial terms in Eq.~\eqref{eq:amp_phi_rbfcoeff} and $d$ denoting the dimensionality of the parameter space (here, $\nu = 6$ and $d = 4$ for the four-dimensional parameter space interpolation). The `imq' RBF kernel is selected, where $\epsilon$ is the shape factor influencing the RBFs' spread at the nodes, determined through (LOOCV) procedure~\cite{hastie2009elements} (see table~\ref{tab:rbf_kernels} for values of $\epsilon$). In the second patch covering ${\chi_{1z} \in [-0.005, \,0.1]}$, the RBF parameters chosen for the first patch are retained, with the exception of a different $\epsilon$ value for this patch. The seismic cutoff frequency ($f_{low}$) is set at $15$ Hz while $f_{start}$ is equal to $10$ Hz.

Once the interpolants for both patches are ready (see Sec.~\ref{sec:interp_generation}), we can use them to generate the interpolated waveforms at the arbitrary query points within the interpolating sample space. Note that the generation of interpolants is a one-time offline process, and it can be completed well in advance before the parameter estimation of the GW event is initiated. While multiple patches are necessary to adequately cover the desired parameter space ranges, the generation of interpolants is highly parallelizable and can benefit from a multicore setup to expedite this stage. In the specific case of this analysis, the generation of interpolants for each patch was accomplished in ${\sim 20}$ minutes using a CPU setup with $32$ cores each. We use the publically available RBF Python package~\cite{RBF_github} to generate RBF interpolants.

To assess the accuracy of the interpolated waveforms, we generated $20$ sets of $10^3$ query points each, randomly distributed in the intrinsic parameter space within the interpolating region (see table~\ref{tab:priordistr}). We then generated both the true and interpolated waveforms at these query points and computed the mismatches between them, assuming a flat Power Spectral Density (PSD). The median mismatch is found to be $\mathcal{O}(10^{-5})$, which shows the high level of accuracy of the meshfree waveforms in approximating the true waveforms across the interpolating sample space

In terms of the computational speed-ups, the meshfree waveforms can be evaluated in $\sim 2.3$ ms in comparison to $\sim 89$ ms (using frequency domain tapered version of \texttt{NRHybSur3dq8} time-domain model) implying a speed-up factor of $\sim 38$. Note that these timings are for the Python implementation of these waveforms. A ``C implementation'' of the true waveforms can be evaluated in $\sim 10$ ms as shown by ~\cite{NRHybSur3dq8} (about $ 4.3 \times$ slower than the meshfree waveform's Python implementation). Since these numbers are also dependent on the hardware configurations of the machines on which these speed tests are performed, a fair comparison of meshfree speed-ups with their C implementation is not possible until we make a C version of meshfree waveforms.
\begin{table}[!hbt]
\centering
\def\arraystretch{1.4}
\begin{tabular}{>{\raggedright\arraybackslash}p{3cm} >{\raggedright\arraybackslash}p{4cm}}
\toprule
 Parameters     &   Injected values \\
\midrule
 $\mathcal{M}$  &   $32.72\: M_{\odot}$ \\
 $q$            &   $1.22$ \\
 $\chi_{eff}$   &   $-0.0091$ \\
 $d_L$          & $2155 \: \text{Mpc}$ \\
 $t_c$          & $1267963151.3$ \\
 $\alpha$       & $0.036\: \text{rad}$ \\
 $\delta$       & $-0.134\: \text{rad}$ \\
 $\iota$        & $0.518\: \text{rad}$ \\
 $\psi$         & $0\: \text{rad}$ \\
\bottomrule
\end{tabular}
\caption{Injected parameters of the simulated BBH event. We are taking $q = m_1/m_2$ where $m_1 \geq m_2$. These injection parameters mimic the inferred source parameters of the \texttt{GW200311\_115853} event detected in the O3 science run of the advanced LIGO and Virgo detectors.}
\label{tab:injected_params}
\end{table}
Finally, to test the accuracy of the meshfree model in the context of parameter estimation, we use $\texttt{SEOBNRv4}\_\texttt{ROM}$~\cite{Boh_2017} waveform model to inject $16$ seconds long simulated BBH event (see Table~\ref{tab:injected_params}) into the Gaussian noise with PSDs taken from {\texttt{aLIGOZeroDetHighPower}}~\cite{aLIGO_ZDHP} for both LIGO-Livingston and LIGO-Hanford detectors, and \texttt{AdvVirgo}~\cite{AdVirgo} for Virgo detector with network matched-filter signal-to-noise ratio (SNR) $\sim 18$. The seismic cutoff frequency ($f_{low}$) is set at $15$ Hz, and the high cutoff frequency is equal to the ringdown frequency of the lowest component masses within the chosen range of chirp masses and mass ratios. A sampling frequency of $2048$ Hz is considered. 
We employ dynesty~\cite{speagle2020dynesty, higson2019dynamic}, sampler, a Python implementation of the Nested sampling algorithm, to sample the posterior distribution. The prior distributions and the boundaries of the parameters (to be estimated) are shown in Table~\ref{tab:priordistr}. We choose the following sampler settings for dynesty: ${\texttt{nLive} = 500}$, ${\texttt{nWalks} = 500}$, ${\texttt{dlogz} = 0.1}$, {\texttt{sample} = ``rwalk''}.  Here, the parameter \texttt{nlive} represents the number of live points, which determines the resolution of the sampled posterior distribution. Smaller values of ${\texttt{nLive}}$ might give rise to a poorly sampled distribution (hence evidence) with much faster convergence. Instead, taking a larger value can give us a finely sampled distribution at the expense of lower convergence speeds. \texttt{walks} specifies the minimum number of steps required before a new live point is proposed, which replaces the live point with the lowest likelihood in the nested sampling. \texttt{sample} indicates the chosen approach for generating samples, and \texttt{dlogz} characterizes the remaining prior volume's contribution to the total evidence. In this PE study, the sampling terminates once \texttt{dlogz} reaches a threshold of $0.1$. For a more comprehensive understanding of dynesty's nested sampling algorithm and its practical implementation, one can refer to the following references ~\cite{speagle2020dynesty, sergey_koposov_2023_7600689}.  
\begin{figure}[hbt]
    \centering
    \includegraphics[width=1\linewidth]{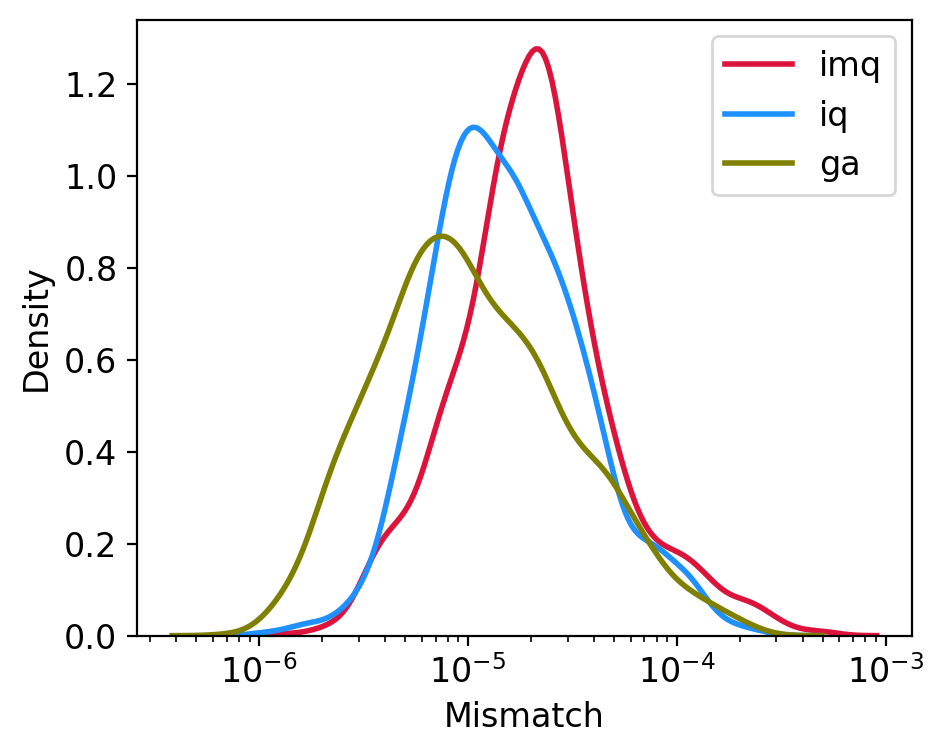}
    \caption{The mismatches calculated between interpolated and true waveforms at the posterior samples are shown. For comparison, we also show the mismatches for meshfree waveforms using `ga' and `iq' RBF kernels. The median mismatched (for all kernels) is $\mathcal{O}(10^{-5})$. Note that the accuracy achieved with `ga' is a little better than with the other kernels due to the high number of nodes that were used to construct `ga' kernel-based meshfree interpolants. However, we do not find any significant difference in the corresponding posterior distributions.}
    \label{fig:mismatches_at_post_samps}
\end{figure}

\begin{figure*}[!hbt]
    \centering
    \includegraphics[width = 0.85\textwidth]{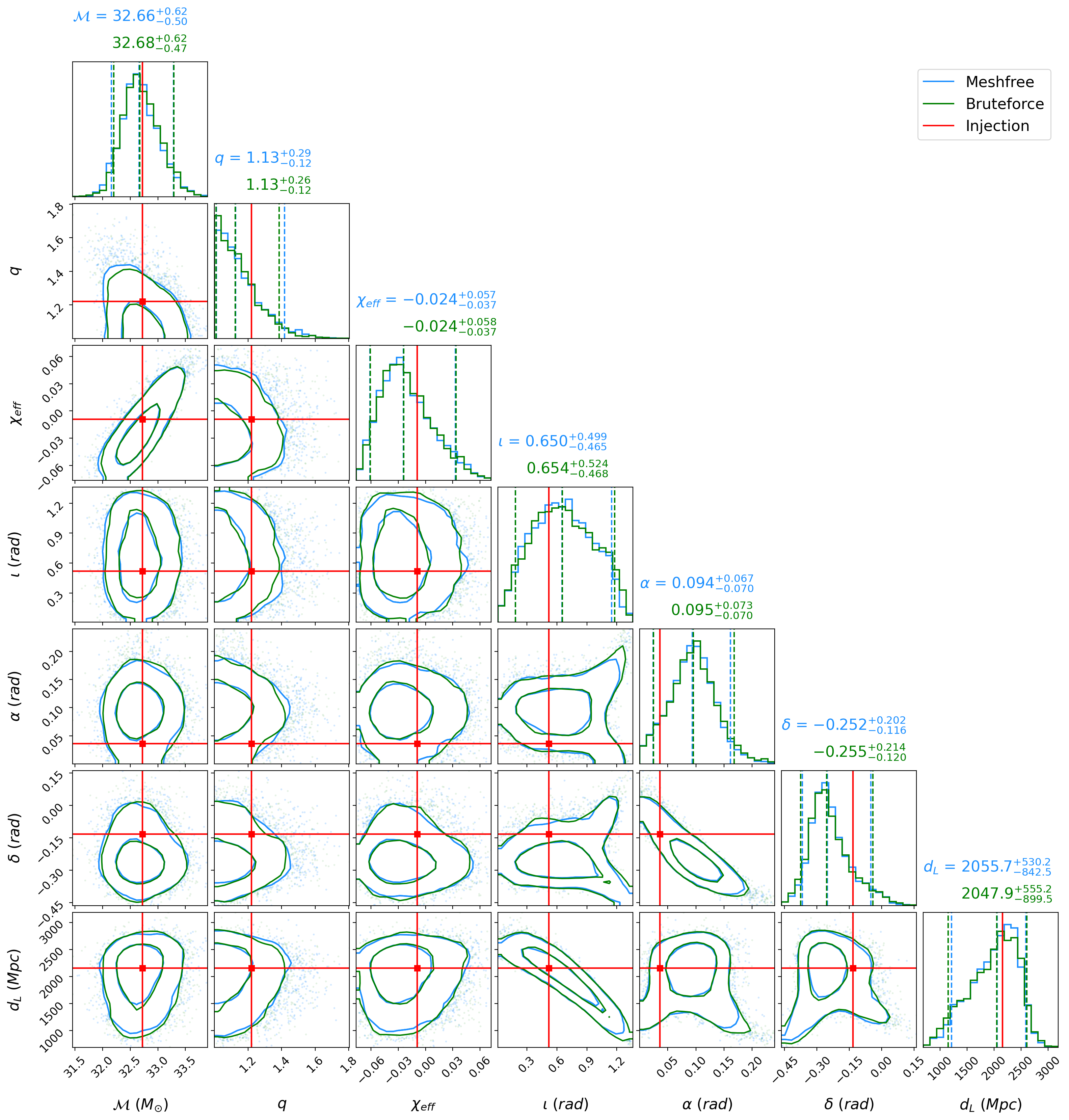}
    \caption{Corner plot of the posterior distributions(with $50\%$ and $90\%$ CI and contours) obtained by the meshfree interpolated waveforms for a simulated BBH event. The vertical red line represents the injected values. It took \mbox{$\sim 16.4$ min} to complete on $64$ CPU cores. In comparison, the PE run with the direct evaluation of the frequency-domain version of the tapered time-domain \texttt{NRHybSur3dq8} waveform model took \mbox{$\sim 5$ h and $50$ min} on the same number of CPU cores and the resulting posterior distributions are broadly consistent with those obtained using meshfree interpolated waveforms.} 
    \label{fig:corner_plot_GW200311_115853}
\end{figure*}

\begin{table}[!hbt]
\def\arraystretch{1.5}
\begin{ruledtabular}
\begin{tabular}{lcl}
 Parameters     &   Range    &   Prior distribution \\
\hline
 $\mathcal{M}$  &   $[25, 40]$ &   $\propto \mathcal{M}$ \\
 $q$            &   $[1, 3]$             & $ \propto \left [ (1 + q)/q^3 \right ]^{2/5}$     \\
 $\chi_{1z}$   &   $[-0.1, 0.1]$   & Uniform  \\
 $\chi_{2z}$   &   $[-0.05, 0.05]$   & Uniform  \\
 $V_{com}$     & $[5e3, 1e11]$                & Uniform\\
 $t_c$          & $t_{\text{trig}} \pm 0.12$& Uniform\\
 $\alpha$       & $[0, 2\pi]$               & Uniform\\
 $\delta$       & $\pm \pi/2$       & $\sin^{-1} \left [ {\text{Uniform}}[-1,1]\right ]$\\
 $\iota$        & $[0, \pi]$                & Uniform in $\cos \iota$\\
 $\psi$         & $[0, 2\pi]$               & Uniform angle\\
\end{tabular}
\end{ruledtabular}
\caption{Prior parameter space over the ten-dimensional parameter space $\vec \Lambda$.}
\label{tab:priordistr}
\end{table}
As evident from Fig.~\ref{fig:corner_plot_GW200311_115853}, posterior distributions of various binary parameters of simulated BBH event contain the injected values well within the  $90\%$ CI. 
\begin{figure}[hbt]
    \centering
    \includegraphics[width=1\linewidth]{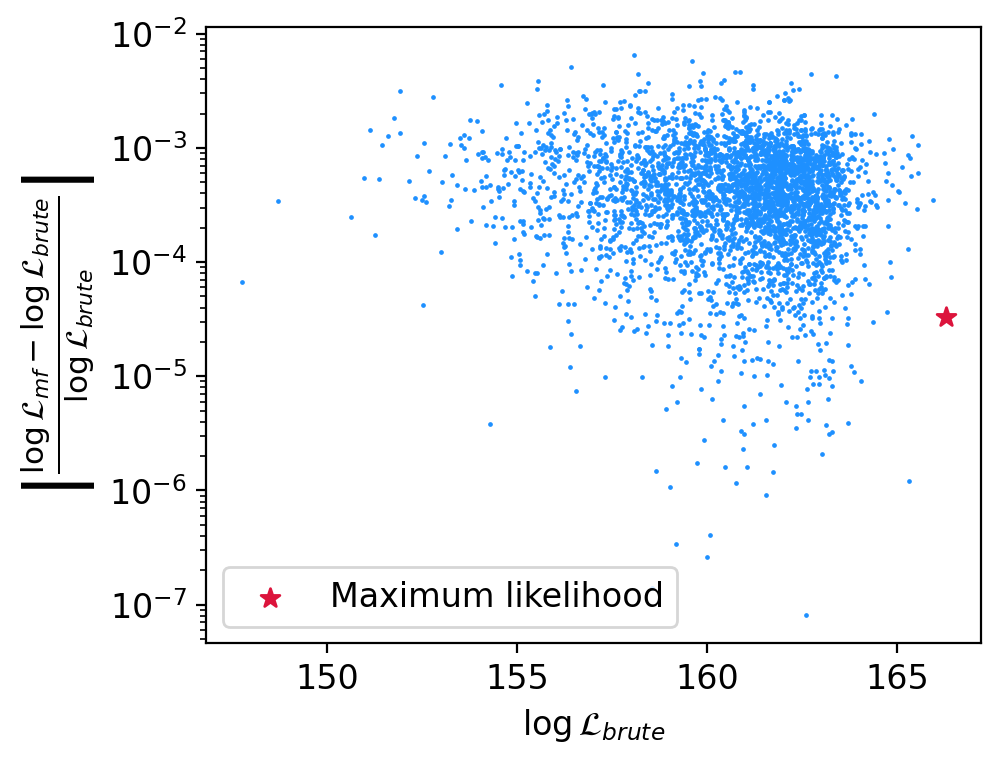}
    \caption{Absolute relative errors in the likelihood as a function of true $\log \mathcal{L}$. The absolute relative error is defined as ${|(\log \mathcal{L}_{\text{mf}} - \log \mathcal{L}_{\text{brute}})/\log \mathcal{L}_{\text{brute}}|}$. The $\log \mathcal{L}_{\text{mf}}$ is calculated using the interpolated waveforms, while $\log \mathcal{L}_{\text{brute}}$ is calculated using the true waveforms. These likelihoods were calculated at the posterior samples obtained from the PE of the simulated BBH event. The red star represents the maximum likelihood posterior sample. Note that the majority of relative errors (especially near the peak) are well within $1\%$, demonstrating the accuracy of the interpolants.}
    \label{fig:like_errors}
\end{figure}
This PE analysis took $\sim 16.4$ minutes on a $64$ CPU cores setup. Fig.~\ref{fig:mismatches_at_post_samps} also shows the probability distribution function (PDF) of the mismatches of the interpolated waveforms with the true waveforms generated at the posterior samples obtained from the PE and median mismatch is $\sim \mathcal{O}(10^{-5})$, demonstrating the good accuracy of the meshfree waveforms.
To quantify the effect of these approximate waveforms on the likelihood calculation, we also evaluated the relative errors in the likelihood, as shown in Fig.~\ref{fig:like_errors} and found a median absolute relative error of $\sim \mathcal{O}(10^{-3})$ further showing the effectiveness of the meshfree waveforms.

All the tests were performed on AMD EPYC $7542$ CPU@$2.90$GHz processors.

\section{Conclusion and Future Outlook}
\label{sec:conclusion}

In this work, we have constructed a fast, high-fidelity 
frequency-domain model for the time-domain \texttt{NRHybSur3dq8} numerical relativity waveform model. We show that such interpolated waveforms can be evaluated in ${\sim 2.3}$ ms, ($\sim 38$) times faster than the frequency-domain tapered version of the time-domain implementation of this waveform model in \textsc{PyCBC}. 
We constructed the waveform interpolants by first selecting a patch of the parameter space. Then, we spray randomly scattered nodes in the sample space and evaluate $\mathcal{A}$ and $\varphi$ from the frequency domain waveform corresponding to \texttt{NRHybSur3dq8} at those nodes. Subsequently, we find a suitable basis spanning the space of both amplitude and phase by performing the SVD of the amplitude and phase matrices resulting from stacking the amplitudes and phases at the nodes.  The resulting SVD coefficients corresponding to amplitude and phase are fit with a linear combination of RBFs and monomials (``interpolants''), followed by the mismatched calculation of the interpolated waveforms with the true waveforms at arbitrary query points in the selected patch of parameter space. The worst-$10$ query points in terms of mismatch are added back into the set of input nodes, and the whole process (called ``Iterative SVD'') repeats till either the maximum mismatch at the given iteration falls below a pre-decided threshold or the number of nodes crosses a fixed number of maximum nodes. In each iteration, the accuracy of the interpolants increases, which in turn enhances the meshfree approximation of the waveforms. Note that the choice of $10$ worst points, as mentioned earlier, is completely arbitrary and chosen to accelerate the interpolants generation process. To test the validity of the meshfree interpolated waveforms, we also performed a parameter estimation of a simulated BBH event using the meshfree interpolated waveforms and found posterior distributions to be consistent with the injected values. It is important to highlight that our method does not require importance sampling as used in deep learning methods such as \textsc{DINGO}~\cite{DINGO} and \textsc{NESSAI}~\cite{NESSAI}.

A limitation we face involves the multiplication of SVD coefficients with the basis vectors. As the duration of the waveform increases, the computational expense of this multiplication rises, consequently impacting the evaluation cost of the mesh-free interpolated waveform. In this context, we can also use the Empirical interpolation method (EIM)~\cite{BARRAULT2004667, Yvon, doi:10.1137/090766498, canizares2015accelerated} to reduce the number of frequency points included in the basis vectors. It is yet to see whether this kind of strategy could be helpful within the meshfree framework in reducing the dimensionality of the basis vectors, which is especially needed for longer-duration waveforms. Further, in this work, we used an iterative SVD scheme that updates the basis vectors with a new set of points by applying independent SVD of the updated amplitude and phase matrix. 
In the follow-up work, we will replace our iterative SVD scheme with advanced updated SVD algorithms ~\cite{BRAND200620, Stange2008OnTE} and will compare the computation cost of performing independent SVD in each iteration and updating the initial basis and singular values from the first iteration using updated SVD algorithms. 
Another notable aspect is that a Euclidean distance measure is employed in Eq.\eqref{eq:amp_phi_rbfcoeff}, even though the parameter space we operate in is inherently non-Euclidean.  Nevertheless, we choose nodes in chirp-time coordinates ($\theta_{0}$ and $\theta_{3}$) instead of component masses or chirp mass and mass ratio as the metric varies slowly in chirp-time coordinates.  Additionally, it's worth noting that this work solely focuses on the aligned spin waveform model \texttt{NRHybSur3dq8}. Subsequent extensions of this work will explore broadening the mesh-free framework to incorporate NR surrogate models encompassing subdominant modes, considering spins, and addressing the aforementioned limitations. We also performed similar exercises for other aligned spin waveform models, such as \texttt{IMRPhenomXAS} and $\texttt{SEOBNRv4}\_\texttt{ROM}$. Since these waveform models are already fast to evaluate, we don't get any significant speed-up in their corresponding meshfree waveform evaluation. However, the speed-up can be further enhanced by employing strategies such as adaptive frequency resolution~\cite{Morisaki:2021ngj} and the empirical interpolation method (EIM) procedure, as mentioned earlier. In future follow-ups of this work, we would also consider extending this mesh-free framework to include NR surrogate models that include subdominant modes, precessing spins, and possibly eccentric models.

\begin{acknowledgements}
 We thank Srashti Goyal for carefully going through the manuscript and giving helpful comments. We especially thank the anonymous referee for their careful review and helpful suggestions. We thank Prayush Kumar for his help in the initial stages of this work. We also thank Abhishek Sharma and Sachin Shukla for their useful suggestions and comments. L.~P. is supported by the Research Scholarship Program of Tata Consultancy Services (TCS). A.~R is supported by the research program of the Netherlands Organisation for Scientific Research (NWO).  A.~S. gratefully acknowledges the generous grant provided by the Department of Science and Technology, India, through the DST-ICPS (Interdisciplinary Cyber Physical Systems) cluster project funding. We thank the HPC support staff at IIT Gandhinagar for their help and cooperation. The authors are grateful for the computational resources provided by the LIGO Laboratory and supported by the National Science Foundation Grants No. PHY-0757058 and No. PHY-0823459. This material is based upon work supported by NSF's LIGO Laboratory, which is a major facility fully funded by the National Science Foundation. 

This research has made use of data or software obtained from the Gravitational Wave Open Science Center~\cite{gwosc_web}, a service of the LIGO Scientific Collaboration, the Virgo Collaboration, and KAGRA. This material is based upon work supported by NSF's LIGO Laboratory, which is a major facility fully funded by the National Science Foundation, as well as the Science and Technology Facilities Council (STFC) of the United Kingdom, the Max-Planck-Society (MPS), and the State of Niedersachsen/Germany for support of the construction of Advanced LIGO and construction and operation of the GEO600 detector. Additional support for Advanced LIGO was provided by the Australian Research Council. Virgo is funded through the European Gravitational Observatory (EGO), the French Centre National de Recherche Scientifique (CNRS), the Italian Istituto Nazionale di Fisica Nucleare (INFN), and the Dutch Nikhef, with contributions by institutions from Belgium, Germany, Greece, Hungary, Ireland, Japan, Monaco, Poland, Portugal, Spain. KAGRA is supported by the Ministry of Education, Culture, Sports, Science and Technology (MEXT), Japan Society for the Promotion of Science (JSPS) in Japan; National Research Foundation (NRF) and Ministry of Science and ICT (MSIT) in Korea; Academia Sinica (AS) and National Science and Technology Council (NSTC) in Taiwan.

\end{acknowledgements}

\bibliography{reference}

\end{document}